\documentclass[12pt]{article}

\usepackage{amsmath, amssymb}
\usepackage{graphicx}
\usepackage{url}
\usepackage{natbib}
\usepackage{hyperref}
\usepackage{colonequals}
\usepackage{enumitem}
\usepackage{float}
\usepackage{xcolor}
\usepackage{bm}
\usepackage{setspace}
\usepackage{multirow}
\usepackage{adjustbox}
\usepackage{booktabs}
\usepackage{ragged2e}
\usepackage{caption}
\usepackage{subcaption}
\usepackage{authblk}
\def\*#1{\mathbf{#1}}
\def\^#1{\amsmathbb{#1}}
\def\##1{\mathbb{#1}}
\DeclareSymbolFontAlphabet{\amsmathbb}{AMSb}%

\begin{document}

\title{Glucodensity Functional Profiles Outperform Traditional Continuous Glucose Monitoring Metrics}

\author[1]{Marcos Matabuena}
\author[2]{Rahul Ghosal}
\author[3]{Javier Enrique Aguilar}
\author[4]{Robert Wagner}
\author[5]{Carmen Fernández Merino}
\author[5]{Juan Sánchez Castro}
\author[6]{Vadim Zipunnikov}
\author[1]{Jukka-Pekka Onnela}
\author[7]{Francisco Gude}

\affil[1]{Department of Biostatistics, Harvard University, USA}
\affil[2]{Department of Epidemiology and Biostatistics, University of South Carolina, USA}
\affil[3]{TU Dortmund}
\affil[4]{Heinrich Heine University Düsseldorf, Department of Endocrinology and Diabetology}
\affil[5]{Primary Care Center, A Estrada, Santiago de Compostela, Spain}
\affil[6]{Department of Biostatistics, Johns Hopkins University, USA}
\affil[7]{Department of Clinical Epidemiology, Complejo Hospitalario Universitario, Santiago de Compostela, Spain;}

\maketitle

\begin{abstract}
Continuous glucose monitoring (CGM) data has revolutionized the management of type 1 diabetes, particularly when integrated with insulin pumps to mitigate clinical events such as hypoglycemia. Recently, there has been growing interest in utilizing CGM devices in clinical studies involving healthy and diabetes populations. However, efficiently exploiting the high temporal resolution of CGM profiles remains a significant challenge. Numerous indices---such as time-in-range metrics and glucose variability measures---have been proposed, but evidence suggests these metrics overlook critical aspects of glucose dynamic homeostasis. As an alternative method, this paper explores the clinical value of glucodensity metrics in capturing glucose dynamics---specifically the speed and acceleration of CGM time series---as new biomarkers for predicting long-term glucose outcomes. Our results demonstrate significant information gains, exceeding 20\% in terms of adjusted $R^2$, in forecasting glycosylated hemoglobin (HbA1c) and fasting plasma glucose (FPG) at five and eight years from baseline AEGIS data, compared to traditional non-CGM and CGM glucose biomarkers. These findings underscore the importance of incorporating more complex CGM functional metrics, such as the glucodensity approach, to fully capture continuous glucose fluctuations across different time-scale resolutions.

\end{abstract}

\section{Introduction}
Recent technological advances in wearable devices and smartphones now enable near-continuous, time-dependent monitoring of various physiological parameters in humans, such as energy expenditure and heart rate \cite{10.1371/journal.pbio.2001402}. Modern medical devices have become a significant driving force in the clinical evolution of digital health and personalized medicine \cite{steinhubl2018digital}. One of the most notable examples of the impact of technological transformation in modern healthcare is the use of continuous glucose monitors (CGMs), a minimally invasive technology that measures interstitial glucose levels every few minutes. Over the past 20 years, CGMs have revolutionized the management of Type 1 diabetes, often integrated with insulin pump systems, significantly reducing high-risk clinical events such as hypoglycemia \cite{Cryer1902}.

Beyond their use in controlling glucose levels and managing Type 1 diabetes, CGMs have introduced new clinical criteria for assessing the effects of new drugs, including next-generation insulins, in both randomized clinical trials and observational studies \cite{gomez2024insulin, battelino2023continuous}. Today, they are widely used to define new clinical outcomes in diabetes research, reshaping how clinical studies in diabetes are conducted. Additionally, CGM devices are essential tools for developing reliable quantitative methods in the field of personalized nutrition \cite{zeevi2015personalized}. For example, recent epidemiological studies have revealed considerable variability in individual glucose responses to the same diets, offering new insights into the effects of meal composition—including macronutrient and micronutrient ratios—on glucose responses over time \cite{matabuena2024multilevel}. CGMs are now being used as a formal criterion for personalizing diets based on individual glycemic responses and for promoting nutritional interventions to optimize metabolic capacity \cite{zeevi2015personalized}
.

Despite strong evidence supporting the benefits of incorporating CGM technology into clinical practice to enhance glycemic control in diabetes, the use of CGMs in healthy, normoglycemic, and prediabetic individuals has not been widely explored or validated \cite{oganesova2024innovative}. However, there is growing interest in using CGMs among healthy populations to estimate diabetes risk and develop interventions aimed at improving metabolic health \cite{10.1371/journal.pbio.2005143}.
Ongoing large-scale studies in the United States and Israel, involving thousands of participants, aim to understand how CGM data can predict diabetes onset, personalize dietary recommendations, and identify early metabolic dysfunctions. These studies seek to fill gaps in the literature regarding the utility of CGMs in non-diabetic populations, providing normative CGM data for glucose variability in healthy individuals \cite{shilo2024continuous}.

While these studies are still in their early stages, the population-based Spanish AEGIS study, with a 10-year follow-up, serves as a key reference for exploring CGM use in healthy populations and obtaining new clinical findings to be validated in larger ongoing cohorts. From the practical perspectives of primary care and preventive medicine, the validation of CGMs in non-diabetic populations is of paramount importance for developing preventive strategies and early interventions to reduce the incidence of diabetes and other metabolic disorders.

From a data analysis perspective, high-resolution CGM time series data present significant challenges, particularly in real-world clinical scenarios where patients are monitored in free-living conditions and for periods of different lengths. Direct analysis of these time series is often impractical, necessitating the development of specific CGM metrics \cite{doi:10.1089/dia.2013.0051,metricas}. Current state-of-the-art CGM metrics rely on compositional data analysis \cite{Dumuid2020}, which defines glucose target ranges and quantifies the proportion of time an individual spends within each range \cite{doi:10.1177/0962280218808819}. Other, simpler methods summarize the time series using means, standard deviations, or other statistical moments \cite{media,beyond2018need, media2, Beckdc181444}
.
Recently, next-generation CGM analyses have been proposed that use functional data analysis \cite{crainiceanu2024functional} to exploit the continuous and functional nature of the data \cite{matabuena2021glucodensities}, as seen in glucodensity representations  and other distributional data metrics \cite{Matabuena2023,Ghosal2023quantile}. While glucodensity offers a continuous characterization of glucose distributions, it may overlook important dynamic aspects such as glucose variability patterns \cite{variabilidadmon}, rapid fluctuations, and the timing of peaks and nadirs \cite{boyne2003timing}—factors critical in understanding glycemic control and predicting complications. Previous studies have shown that glucose variability and excursions are associated with oxidative stress and vascular complications. Despite these limitations, glucodensity remains attractive due to its ability to provide a comprehensive and interpretable summary of glucose data, incorporating traditional CGM metrics automatically into the glucodensity representation.

Glucodensity \cite{matabuena2021glucodensities} extends traditional compositional metrics like "time in range" by providing a continuous characterization of the proportion of time spent above or below each glucose concentration. Practically, glucodensity transforms the glucose time series into a marginal density function, automatically incorporating traditional CGM metrics. In some clinical applications, glucodensity has proven valuable in predicting clinical outcomes while maintaining interpretability—an essential feature in clinical research. However, new methods based on neural networks, such as autoencoders and other deep learning techniques, have been proposed \cite{lutsker2024glucose}
 to summarize CGM data and integrate it with other data modalities. While these methods can capture complex temporal patterns and glucose dynamics, they often lack interpretability, making it difficult for clinicians to understand and explain the clinical results. In addition, neural network methods require large sample sizes and extensive computational resources, which may not be practical in clinical studies with small or moderate sample sizes, such as clinical trials. In contrast, glucodensity offers a balance between capturing essential glucose information and maintaining interpretability, making it a practical choice for small participant clinical studies.

The main goal of this paper is to explore, for the first time, new characteristics of glucose dynamics over time by focusing on the clinical value of glucose speed (rate of change) and acceleration (rate of change of the rate of change) in order to predict long-term outcomes using the glucodensity approach. While previous studies have attempted to capture glucose variability and dynamics through measures like standard deviation or mean amplitude of glycemic excursions (MAGE) \cite{variabilidadmon, doi:10.1089/dia.2005.7.253}, these methods do not fully capture the continuous and temporal aspects of glucose fluctuations \cite{Zaccardi2018}. By incorporating glucose speed and acceleration into the glucodensity framework, we aim to create a unified functional profile that captures both the distributional and dynamic characteristics of glucose time series. For this analytical purpose, we utilize baseline data from the AEGIS study, a cohort primarily composed of healthy individuals—a random sample from the general population. We focus on predicting key biomarkers for diabetes control and progression—fasting plasma glucose (FPG) and glycosylated hemoglobin (HbA1c)—at two future time points, five and eight years. This approach addresses a critical gap in existing methods to capture glucose variability dynamics from an accuracy perspective.

In our analysis, we employ a glucodensity-based methodology and introduce the novel concept of multivariate glucodensity to simultaneously capture the effects of glucose, glucose speed, and glucose acceleration in a unified functional profile. This multivariate representation is constructed by extending the glucodensity framework to include not only the distribution of glucose concentrations but also the distributions of their first and second derivatives over time, effectively capturing the dynamics of glucose changes.

From a clinical perspective—the primary goal of the paper—we provide solid evidence that glucose speed and acceleration over time are promising biomarkers that serve as surrogate measures for glucose control, offering a new clinical outcome measure to comprehensively encapsulate CGM information. The implications of these findings in translational research could have substantial consequences in drug development. For instance, new insulins could be designed to respond more dynamically to rapid changes in glucose levels, optimizing dosing regimens based on individual glucose kinetics rather than static glucose concentrations. This could lead to more effective treatment schedules that minimize glycemic excursions and reduce the risk of complications. From a methodological standpoint, our work opens up new opportunities to define functional clinical biomarkers for diabetes research and practice using advanced CGM metrics that capture both the distributional and dynamic aspects of glucose homeostasis metabolism.

\subsection{Scientific Goals and Contributions}

\begin{itemize}
    \item We introduce novel CGM data analysis methods based on glucodensity approach and distributional data analysis to capture glucose dynamics at different time scales, focusing on the speed and acceleration of glucose levels over time.
    
    \item We validate the proposed CGM metrics by predicting long-term glucose evolution at five and eight years, considering the biomarkers glycosylated hemoglobin (HbA1c) and fasting plasma glucose (FPG)—the primary biomarkers for diabetes control and diagnosis.
    
    \item We employ functional data analysis methods that efficiently leverage CGM data and distributional representations. This approach allows us to capture glucose dynamics in terms of speed and acceleration, translating into greater accuracy in predicting clinical outcomes.
    
    \item Our results show a 20\% increase in $R^2$ for predicting the continuous outcomes FPG and HbA1c at different temporal points when incorporating glucose marginal distributional patterns, speed, and acceleration as predictors in regression models, compared to regression models that only involve traditional CGM and non-CGM metrics. The traditional CGM and non-CGM models include: (i) baseline FPG and HbA1c; and (ii) baseline FPG and HbA1c, along with CGM metrics like the Area Under the Curve (AUC), Mean Amplitude of Glycemic Excursions (MAGE), Continuous Overall Net Glycemic Action (CONGA), and hyperglycemia time-in-range.
\end{itemize}

\section{CGM Data Analysis Methods}

In this section, we introduce the formal framework for analyzing continuous glucose monitor (CGM) data using the glucodensity (distributional data analysis) approach. This method incorporates the dynamics of glucose in terms of speed and acceleration, framed within the context of distributional glucose representations. We begin by presenting the existing concept of the distributional representation of glucose data.

Next, we extend this concept by proposing a novel multidimensional glucodensity approach that aggregates multiple physiological signals, specifically incorporating the speed (first derivative) and acceleration (second derivative) of the continuous-time glucose process. We then discuss the formal methods used to integrate these derivatives into the glucodensity representations, employing smoothing techniques for accurate estimation.

Finally, we introduce specific additive semi-parametric regression models that accommodate both marginal and multivariate distributional representations to predict continuous glucose scalar outcomes.

\subsection{Distributional representations for CGM-data analysis}\label{sec:gluco}
\subsubsection{Univariate Glucodensity approach}
Following \cite{matabuena2021glucodensities},
for a individual $i^{th}$, denote the glucose monitoring data by pairs $(t_{ij}, G_{ij})$, $j = 1,\ldots,n_i,$ where the $t_{ij}$ represent recording times that are typically equally spaced across the observation interval, and $G_{ij}$ is the glucose level at time $t_{ij} \in \mathcal{S}_{i} = [0, T_i].$ Note that the number of records $n_i$, the spacing between them, and the overall observation length $T_i$ can vary by individual. One can think of these data as discrete observations of a continuous latent process $Y_i(t),$ with $G_{ij} = Y_i(t_{ij}).$ The glucodensity for this patient is defined in terms of this latent process as $f_i(x) = F_i'(x),$ where

\begin{align} F_i(x) = \frac{1}{T_i}\int_0^{T_i} \mathbf{I}\left(Y_i(t) \leq x\right) \mathrm{d}t \quad \text{for} \quad \inf_{t \in [0,T_i]} Y_i(t) \leq x \leq \sup_{t \in [0,T_i]} Y_i(t) \end{align}

is the proportion of the observation interval in which the glucose levels remain below $x.$ Since $F_i$ is increasing from 0 to 1, the data to be modeled are a set of probability density functions $f_i,$ $i = 1,\ldots,n.$

Of course, neither $F_i$ nor the glucodensity $f_i$ is directly observed in practice. However, one can construct an approximation using a density estimate $\widehat{f}_{i}(\cdot)$ obtained from the observed sample. In the case of CGM data, the glucodensities may have different supports and shapes. Therefore, we suggest using a non-parametric approach, such as a kernel-type estimator, to estimate each density function. Alternatively, for patient $i$, we can use the quantile function $\widehat{\mathcal{Q}}_i(p) = \inf\{s \in \mathbb{R} : \widehat{F}_i(s) \geq p\}$ as a representation, where $\widehat{F}_{i}(s) = \frac{1}{n_i} \sum_{j=1}^{n_i} \mathbf{I}(G_{ij} \leq s)$. 

\begin{figure}[t]
\centering
\begin{subfigure} {0.8\linewidth}
\includegraphics[width=\linewidth]{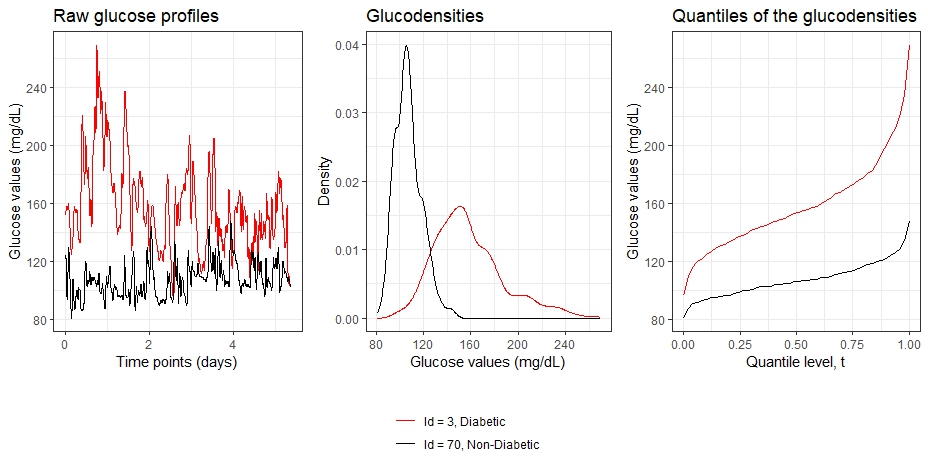}
            \caption{\footnotesize Glucodensity profiles from raw CGM data for a diabetic and non diabetic individual.}
            \label{fig:glucodensities_1}
            \end{subfigure}
\end{figure}

Figure \ref{fig:glucodensities_1} illustrates graphically the process of collapsing CGM information from the raw CGM series into the quantile and density distributions. Density functions offer more interpretability than quantile functions because they provide a surrogate measure for the distribution of individuals across each glucose concentration. In contrast, quantile functions represent the profile of an individual patient. They are linear statistical objects that take values in vector-valued spaces and can be analyzed using standard linear functional data methods.

In metabolic disorders, such as diabetes, it is biologically plausible to assume that disease  remains stable over certain periods. Consequently, we can hypothesize that physiological responses, including short-term glucose variation during these periods, conform to a stationary distribution, specifically over intervals $\mathcal{S}_{i}$. Therefore, the glucodensity approach is well-suited to capture individual distributional patterns within the examined temporal periods of glucose values.

%\subsubsection{Glucodensity  multivariate approach}\label{sec:mult}

\subsection{Multivariate Glucodensity Approach}

\begin{figure}[H]
    \centering
    \includegraphics[width=0.9\textwidth]{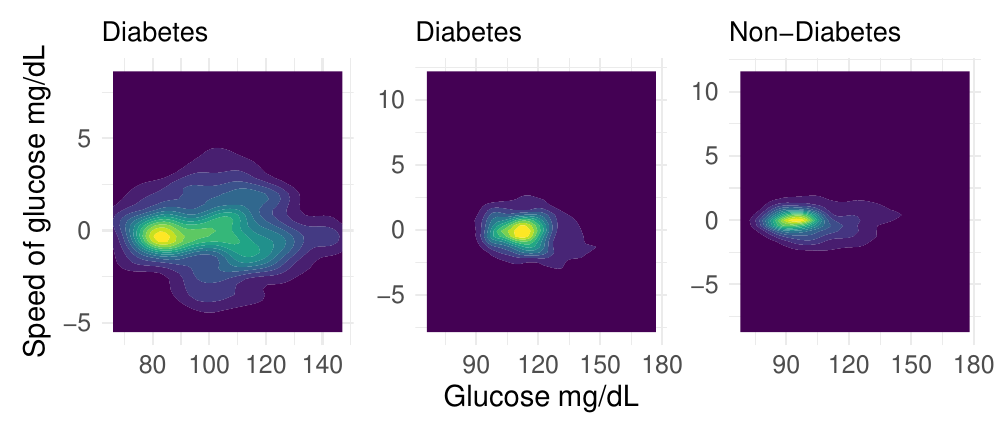}  % Cambia esta línea con la ruta a tu imagen
    \caption{Heatmap of the two-dimensional density for glucose concentration and its first derivative (speed) estimated  from the CGM time series for three non-diabetic individuals at baseline of the study. After 8 years, the individuals in the left and middle panels developed diabetes, while the individual in the right panel did not.}
    \label{fig:densmulti}
\end{figure}

We now extend the notion of glucodensity to cases where multiple physiological parameters are observed simultaneously—the multidimensional case. Figure \ref{fig:densmulti} shows the heatmap estimation for three non-diabetic individuals, representing two-dimensional density of both glucose and and its first derivative (speed). This provides a clear intuition that, to capture glucose fluctuations at different time scales and measure local variability patterns, more complex CGM representations are necessary. Notably, in the individual whose final status remains non-diabetic (right panel), there is a more concentrated density within the glucose range considered normal and less variability in glucose speed compared to the two individuals who developed diabetes after a certain time. These descriptive  plots suggest that the dynamics of CGM could contain indicative information in measuring the risk and progression of Diabetes Mellitus.

For a individual $i$, we observe $n_i$ measurements over the time interval $\mathcal{S}_{i} = [0, T_i]$ from a medical device at the temporal instants $\Gamma_i = \{t_{ij}\}_{j=1}^{n_i} \subset \mathcal{S}_i$ for $m$ physiological variables recorded simultaneously at the same sampling frequency. The measurements are denoted as $G_{ijk} = Y_{ik}(t_{ij})$, where $Y_{ik}(\cdot)$ is the underlying continuous model, and $k = 1, \dots, m$, $i = 1, \dots, n$, $j = 1, \dots, n_i$.

Throughout this paper, we consider $m = 3$ physiological variables: glucose concentration (measured in mg/dL) and its first and second derivatives with respect to time, representing the speed and acceleration of glucose change. In the AEGIS study, the CGM device is continuously worn until the end of the monitoring period; thus, there is no missing data in the CGM time series collected. We then smooth the raw functional data trajectories via $B$-spline basis expansion and define
\[
G_{ij1} = \widehat{Y}_{i1}(t_{ij}),
\]
where
\[
\widehat{Y}_{i1}(t_{ij}) = \sum_{l=1}^{r_i} c_{il} \phi_{l}(t_{ij}), \quad i = 1, \dots, n, \; j = 1, \dots, n_i, \; l = 1, \dots, r_i,
\]
$\phi_{l}(t_{ij})$ are the basis functions predefined at the beginning of functional data smoothing, and $c_{il}$ are the individual coefficients. The smoothing criterion consists of minimizing the mean squared error while introducing a quadratic penalty term to control the level of smoothness.

Formally, we define the sequences $G_{ij2}$ and $G_{ij3}$ as the evaluations at times $t_{ij}$ of the first and second derivatives of the smoothing representation $\widehat{Y}_{i1}(t_{ij})$, that is,
\[
G_{ij2} = \widehat{Y}'_{i1}(t_{ij}) \quad \text{and} \quad G_{ij3} = \widehat{Y}''_{i1}(t_{ij}).
\]

From a population standpoint, we define the cumulative distribution function (c.d.f.) of the multivariate process $Y_i(t) = (Y_{i1}(t), \dots, Y_{im}(t))$ over the interval $[0, T_i]$ as $F_i(p_1, \dots, p_m)$, described by the following equation:
\[
F_i(p_1, \dots, p_m) = \frac{1}{T_i} \int_{0}^{T_i} \mathbf{I}\{Y_i(t) \leq \mathbf{p}\} \, dt.
\]

The corresponding individual density function $f_i(p_1, \dots, p_m)$ is defined as the gradient of $F_i(p_1, \dots, p_m)$:
\[
f_i(p_1, \dots, p_m) = \nabla F_i(p_1, \dots, p_m).
\]

From an empirical perspective, assuming the original process $Y_{ik}(t)$ is smooth, we estimate the underlying marginal density functions using smoothing methods as follows:
\[
\widehat{f}_{\mathbf{H}}(\mathbf{p}) = \frac{1}{n} \sum_{i=1}^n K_{\mathbf{H}} (\mathbf{p} - \mathbf{p}_i),
\]
where $\mathbf{p} = (p_1, p_2, \dots, p_m)^\top$, $\mathbf{p}_i = (p_{i1}, p_{i2}, \dots, p_{im})^\top$ for $i = 1, 2, \dots, n$ are $m$-vectors; $\mathbf{H}$ is a symmetric and positive definite $m \times m$ matrix that serves as the bandwidth; and $K$ is the kernel function, a symmetric multivariate density defined as:
\[
K_{\mathbf{H}}(\mathbf{p}) = |\mathbf{H}|^{-1/2} K(\mathbf{H}^{-1/2} \mathbf{p}).
\]

To select the bandwidth matrix $\mathbf{H}$, we can use rules of thumb based on asymptotic Gaussian processes or finite-sample data-driven approaches such as cross-validation (see more details in \cite{chacon2018multivariate}).

\subsection{Regression modeling}\label{sec:reg}
\subsubsection{Modelling framework}
%\textcolor{red}{Rahul Rewrite}
We denote the scalar outcome of interest as $R_i$. Let us denote the scalar covariates e.g., age and other demographic predictors as $\*X_i$. The subject-specific marginal distributional representation using quantile function of glucose concentration, speed and acceleration is denoted by $\mathcal{Q}_{i1}(\cdot),\mathcal{Q}_{i2}(\cdot)$ and $\mathcal{Q}_{i3}(\cdot)$ respectively. The quantile function representation has been previously used in distributional data analysis \citep{Ghosal2023quantile} and offers attractive mathematical advantages. We observe a random sample $\*D_i=(R_i, \*X_i, \mathcal{Q}_{i1}(\cdot),\mathcal{Q}_{i2}(\cdot),\mathcal{Q}_{i3}(\cdot))$, $i=1,\dots, n$, which are assumed to be independent and identically distributed for each $i$.

\subsection{Scalar on Distribution Regression}\label{sec:res}
 We use the following additive scalar on distribution regression approach \citep{Ghosal2023quantile,Matabuena2023} for modelling the scalar outcome of interest based on multivariate distributional representations. A functional generalized additive model \citep{mclean2014functional} is used for capturing additive nonlinear distributional effects of glucose concentration, speed and acceleration. The model is given by,
\begin{equation}
    R_i =  \alpha_0+\*X_i^T\bm\alpha + \int_{P_1} F_{1}( \mathcal{Q}_{i1}(p_1),p_1)dp_1+\int_{P_2} F_{2}(\mathcal{Q}_{i2}(p_2),p_2) dp_2+\int_{P_3} F_{3}(\mathcal{Q}_{i3}(p_3),p_3) dp_3+ \epsilon_i, 
\end{equation}

\noindent where $F_{j}(\cdot,\cdot),j=1,\dots,3$ denote a unknown bivariate function capturing the additive effect of ${Q}_{ij}(p_j)$ at index $p_j$. The unknown parameters of interest in the above model, which needs to be estimated are given by $\bm\Theta=(\alpha_0,\bm\alpha,F_{1}(\cdot,\cdot),F_{2}(\cdot,\cdot),F_{3}(\cdot,\cdot))$. Note that we do not need directly observed these distribution valued covariates, rather they are empirically estimated.

\subsection{Estimation}
We employ a semiparametric penalized estimation approach to estimate the model parameters.
The unknown bivariate functional effects $F_{j}(\cdot,\cdot)$ (j=1,\dots,3) are modelled using tensor 
product of cubic B-spline basis functions as
\begin{equation}
 F_{j}(u,p)= \sum_{k=1}^{K_0} \sum_{l=1}^{L_0} \theta_{j,kl} B_{j,U}(u) B_{j,\mathcal{P}}(p).
\end{equation}
Here $\{B_{j,U}(u) \}_{k=1}^{K_0}$, $\{B_{j,\mathcal{P}}(p)\}_{l=1}^{L_0}$ are sets of known basis functions over $u$ and $p$ arguments respectively. We denote $\bm\theta_j=\{\theta_{j,kl}\}_{k=1,l=1}^{K_0,L_0}$ to be the unknown basis coefficients. In this article, we use cubic B-spline basis functions, however, other basis functions can be used as well. Plugging in these basis expansions in model (3) we have,

\begin{equation}
\begin{aligned}
 R_i &=  \alpha_0+ \*X_i^T\bm\alpha + \sum_{j=1}^3\int_{P_j} \sum_{k=1}^{K_0} \sum_{l=1}^{L_0} \theta_{j,kl} B_{j,U}(\mathcal{Q}_{ij}(p_j)) B_{j,P}(p_j)dp_j.  \\
&=  \alpha_0+ \*X_i^T\bm\alpha +  \sum_{j=1}^3 \sum_{k=1}^{K_0} \sum_{l=1}^{L_0} \theta_{j,kl}  \int_{P_j} B_{j,U}(\mathcal{Q}_{ij}(p_j)) B_{j,P}(p_j)dp_j.  \\
&= \alpha_0+ X_i^T\bm\alpha +  \sum_{j=1}^3  \*W_{ij}^T \bm\theta_j.
\end{aligned}
\end{equation}
Here $\hat{W}_{ij}=\{\int_{P_j} B_{j,U}(\mathcal{Q}_{ij}(p_j)) B_{j,P}(p_j)dp_j\}_{k=1,\ell=1}^{K_0,L_0}$ denotes the $K_0L_0$-dimensional stacked vectors which can be approximated using Riemann sum. Denote the  unknown parameter of interest as $\bm\psi=(\alpha_0,\bm\alpha^T,\bm\theta_1^T,\bm\theta_2^T,\bm\theta_3^T)^T$. We use the following penalized least square criterion to estimate the model parameters which simultaneously estimate the basis coefficients an enforces smoothness in the additive functional effects $F_{j}(\cdot,\cdot)$ ($j=1,\ldots,3$).
\begin{eqnarray}
    R_{p}(\bm\psi|\bm\lambda) = R(\bm\psi|\bm\lambda)+\sum_{j=1}^3 \bm\theta_j^T \*P_{j} \bm\theta_j
\notag \\ 
=\sum_{i=1}^{n} (R_i-\alpha_0- \*X_i^T\bm\alpha -  \sum_{j=1}^3  \hat{W}_{ij}^T \bm\theta_j)^2+\sum_{j=1}^3 \bm\theta_j^T \hat{P}_{j} \bm\theta_j, \label{ppl}
\end{eqnarray}
where is a $\*P_1$ and $\*P_2$ are roughness penalty matrices \citep{happ2018} to introduce smoothness in $u$ and $p$ direction respectively for each of the function $F_{j}(\cdot,\cdot)$. In this article, we have used a second order difference penalty  \citep{marx2005multidimensional} 
 which introduces smoothness in in both arguments $u$ and $p$. The penalty matrices $\^P_j$ ($j=1\ldots 3$) are given by $\hat{P}_j=\lambda_{U,j}\hat{D}_{U,j}^T\hat{D}_{U,j} \otimes\hat{I}_{L_0}+\lambda_{P,j} \hat{I}_{K_0}\otimes\hat{D}_{P,j}^T\hat{D}_{P,j} $, where $\hat{I}_{K_0},\*I_{L_0}$ are identity matrices with dimension $K_0,L_0$ and $\hat{D}_{U,j}$ , $\hat{D}_{P,j}$ are matrices corresponding to the row and column of second order difference penalties. The penalty parameters $\lambda_{U,j},\lambda_{P,j}$ ($j=1\ldots 3$) are the corresponding smoothing parameters, controlling smoothness of $F_{j}(\cdot,\cdot)$ over $u$ and $p$ respectively. For fixed values of $\bm\lambda$, Newton-Raphson algorithm can be employed for estimating $\bm\psi$ as the maximizer of the above penalized least square. We used the \texttt{gam} function within \texttt{mgcv} package in R \citep{wood2015package} for estimating the parameter vector $\bm\psi$. The smoothing parameters $\bm\lambda$ can be chosen based on data driven criterion like generalized cross validation (GCV) \citep{ruppert2003semiparametric} or restricted maximum likelihood criteria (REML) \citep{kim2016general}.

\section{AEGIS study}
The AEGIS population study \cite{Gude2017} is a ten-year longitudinal study that focuses on changes in circulating glucose and its connections to inflammation and obesity. These factors are critically linked to the potential development of comorbidities and diabetes mellitus. Unlike other epidemiological studies, AEGIS incorporates continuous glucose monitoring (CGM) for a random subsample, providing detailed glucose profiles at various time points over a period of five years. In this random AEGIS CGM subsample, most individuals are non-diabetic (normoglycemic and prediabetic), offering a unique opportunity to evaluate the clinical value of CGM in healthy populations. The diabetic individuals in the sample have Type II diabetes and are very well-controlled, with a  minimal progression of the disease at baseline. For this reason, we refer  that with analysis,  we are characterize the long-term risk and progression of Diabetes Mellits disease.

\subsection{Sample and Procedures}
\label{sec:material}
\subsubsection{Study design}

A subset of the subjects in the A Estrada Glycation and Inflammation Study (AEGIS; trial $NCT01796184$ at \url{www.clinicaltrials.gov}) provided the sample for the present work. In the latter cross-sectional study, an age-stratified random sample of the population (aged $\geq 18)$ was drawn from Spain's National Health System Registry. A detailed description has been published elsewhere \citep{Gude2017}. For a one-year period beginning in March, subjects were periodically examined at their primary care center where they $(i)$ completed an interviewer-administered structured questionnaire; $(ii)$ provided a lifestyle description; $(iii)$ were  to biochemical measurements, and $(iv)$ were prepared for CGM (lasting $6$ days). The subjects who made up the present sample were the $581$ ($361$ women, $220$ men) who completed at least $2$ days of monitoring, out of an $622$  persons who consented to undergo a $6$-day period of CGM.  Another $41$ original subjects were withdrawn from the study due to non-compliance with protocol (n = $4$) or difficulties in handling the device (n = $37$). The characteristics of the participants are shown in the Table  \ref{table:tabla4}.

\begin{table}[ht]
	\centering
	\scalebox{1}{
	\begin{tabular}{lll}
		\hline
		& Men $(n=220)$ & Women $(n=361)$ \\ 
		\hline
		Age, years & $47.8\pm 14.8$ & $48.2\pm14.5$ \\ 
		A1c, \% & $5.6\pm0.9$ & $5.5\pm0.7$ \\ 
		FPG $mg/dL$ & $97\pm23$ & $91\pm21$ \\ 
		HOMA-IR $mg/dL.\mu UI/m$ & $3.97\pm5.56$ & $2.74\pm2.47$ \\ 
		BMI $kg/m^2$ & $28.9\pm4.7$ & $27.7\pm5.3$ \\ 
		CONGA $mg/dL$  & $0.88\pm0.40$ & $0.86\pm0.36$ \\ 
		MAGE $mg/dL$ & $33.6\pm22.3$ & $31.2 \pm14.6$ \\ 
		MODD & $0.84\pm0.58$ & $0.77\pm0.33$ \\ 
		\hline
	\end{tabular}} 
	\caption{Characteristics of AEGIS study participants by sex. Mean $\pm$ standard deviation are shown.	$BMI$ - body mass index; $FPG$ - fasting plasma glucose; $A1c$ - glycated haemoglobin; $HOMA-IR$ - homeostasis model assessment-insulin resistance; $CONGA$ - glycemic variability in terms of continuous overall net glycemic action; $MODD$ - mean of daily differences; $MAGE$ - mean amplitude of glycemic excursions.}
	\label{table:tabla4}
\end{table}

\subsubsection{Laboratory determinations}

Glucose was determined in plasma samples from fasting participants by the glucose oxidase peroxidase method. A1c was determined by high-performance liquid chromatography in a Menarini Diagnostics HA-$8160$ analyzer; all A1c values were converted to DCCT-aligned values \cite{Hoelzel166}. 
Insulin resistance was estimated using the homeostasis model assessment method (HOMA-IR) as the fasting concentration of plasma insulin ($\mu$ units/mL) $\times$  plasma glucose (mg/dL)/ $405$.

\subsubsection{CGM Procedures}

At the start of each monitoring period, a research nurse inserted a sensor (Enlite™, Medtronic, Inc, Northridge, CA, USA) subcutaneously into the subject's abdomen and instructed him/her in the use of the iPro™ CGM device (Medtronic, Inc, Northridge, CA, USA). The sensor continuously measures the interstitial glucose level $40-400$ (range mg/dL) of the subcutaneous tissue, recording values every $5$ min. Participants were also provided with a conventional OneTouchR VerioR Pro glucometer (LifeScan, Milpitas, CA, USA) as well as compatible lancets and test strips for calibrating the CGM. 

Participants were also provided with a conventional OneTouchR VerioR Pro glucometer (LifeScan, Milpitas, CA, USA) and compatible lancets and test strips. To guarantee the reliability and quality of the monitoring data, the participants were instructed to perform at least 3 capillary blood glucose measurements per day (usually before the main meals). The capillary blood glucose readings were used to calibrate the $iPro^{TM}$ CGM system. Data from monitoring that could not be calibrated with at least 3 capillary blood glucose controls per day were excluded from the analysis.

\subsubsection{Glycaemic variability}
After downloading the recorded data, the following glycemic variability indexes were analyzed: area under the curve (AUC), mean amplitude of glycemic excursions (MAGE) \cite{doi:10.1089/dia.2005.7.253}, continuous overlapping net glycemic action at 1 h (CONGA1) \cite{MAGE}, the mean of daily differences (MODD) \cite{molnar1972day}, and the time above range (TAR). The mathematical formulae of the methods of assessment for glucose variability were taken from their original publications for inclusion in an R program, which is freely available \cite{Gude2017}.

\subsubsection{Ethical approval and informed consent}

The present study was reviewed and approved by the Clinical Research Ethics Committee from Galicia, Spain (CEIC$2012$-$025$). Written informed consent was obtained from each participant in the study, which conformed to the current Helsinki Declaration.

 \section{Results}

\begin{table}[H]
\centering
\caption{Adjusted $R^2$ summary for clinical outcomes and the five predictive models of varying complexity. Model (1) includes baseline age, FPG, and HbA1c; Model (2) adds CGM metrics such as Area Under the Curve (AUC), Mean Amplitude of Glycemic Excursions (MAGE), Continuous Overall Net Glycemic Action (CONGA), and hyperglycemia time-in-range; Model (3) adds the quantile function for the glucose profile; Model (4) adds quantile functions for glucose speed; and Model (5) adds quantile functions for glucose acceleration.}
\label{tab:adjusted_r_square_summary}
\begin{tabular}{lccccc}
\toprule
{} &  Model (1) &  Model (2) &  Model (3) &  Model (4) &  Model (5) \\
\midrule
HbA1c--5 years &      0.640 &      0.645 &      0.732 &      0.743 &      0.796 \\
HbA1c--8 years &      0.584 &      0.579 &      0.628 &      0.656 &      0.680 \\
FPG--5 years   &      0.434 &      0.438 &      0.518 &      0.587 &      0.617 \\
FPG--8 years   &      0.339 &      0.426 &      0.425 &      0.445 &      0.517 \\
\bottomrule
\end{tabular}
\end{table}
The primary objective of the Results section is to provide evidence of the predictive superiority of the CGM distributional metrics introduced in Section \ref{sec:gluco}, which incorporate glucose speed and acceleration to predict clinical outcomes. To evaluate these metrics, we compare models of varying complexity that include both CGM and non-CGM variables from the AEGIS baseline data, using the regression methodology outlined in Section \ref{sec:reg}. Given the limited sample size and the presence of missing data in the five- and eight-year outcomes analyzed, which further reduces the final sample size, we focus on introducing glucose speed and acceleration in a marginal yet additive manner within the regression models. For each clinical outcome, we fit the model independently using the available individual data, as the number of missing responses varies across the examined variables. The baseline data do not have any missingness.

We assess the predictive performance of these models using adjusted $R^2$, which allows us to fairly compare models while accounting for complexity and avoiding overfitting. The CGM distributional metrics, including glucose speed and acceleration, are compared with traditional CGM and non-CGM biomarkers.

Table \ref{table:tabla4} displays the baseline patient characteristics from the AEGIS population sample by sex, indicating that the sample is primarily composed of non-diabetic individuals. In cases where patients have diabetes, they are very well-controlled, with glucose values very close to the prediabetes range. For the characterization of baseline data, focus on monitors the risk and progression of diabetes in a primarily healthy cohort. Among individuals with diabetes, the progression of the disease remains minimal.

We consider four clinical outcomes: (i) HbA1c at 5 years, (ii) HbA1c at 8 years, (iii) FPG at 5 years, and (iv) FPG at 8 years. Five models based on baseline data of increasing complexity are evaluated:

\begin{itemize}
    \item \textbf{Model (1):} Includes age, FPG, and HbA1c.
    \item \textbf{Model (2):} Adds traditional CGM metrics, such as MAGE, CONGA, and Hyper, to age, FPG, and HbA1c.
    \item \textbf{Model (3):} Incorporates age, FPG, HbA1c, and the glucodensity quantile marginal distributional representation.
    \item \textbf{Model (4):} Extends Model (3) by including quantiles of glucose speed.
    \item \textbf{Model (5):} Extends Model (4) by adding quantiles of glucose acceleration.
\end{itemize}

To predict the clinical outcomes, denoted as $R^j$ where $j \in \{\text{HbA1c--5, HbA1c--8, FPG--5, FPG--8}\}$, we employ semi-parametric additive generalized models, as described in Section \ref{sec:mult}. The univariate quantile functions for glucose, speed, and acceleration are estimated on a grid of 100 points, following the smoothing methodology from Section \ref{sec:mult}. Each individual distributional-quantile representation $\mathcal{Q}_{fij}$, $j \in \{1, 2, 3\}$, is modeled using a bivariate spline approach outlined in McLean et al. (2014) \cite{mclean2014functional}. Continuous scalar variables are incorporated into the regression model as linear terms.

Table \ref{tab:adjusted_r_square_summary} presents the predictive performance of the five models across the four clinical outcomes, using adjusted $R^2$. This metric accounts for the number of parameters, allowing for a balanced comparison across models with different complexities.

Model (5), which incorporates glucose, glucose speed, and acceleration, consistently outperforms the other models, demonstrating significant information gains compared to Models (1) and (2). For example, Model (5) shows a 23\% increase in adjusted $R^2$ for HbA1c at 5 years compared to Model (2), and an 8.7\% increase compared to Model (3). For HbA1c at 8 years, Model (5) achieves a 17.4\% improvement over Model (2) and an 8.3\% increase compared to Model (3). The gains are even more pronounced for fasting plasma glucose (FPG) predictions: at 5 years, Model (5) shows a 40.9\% improvement over Model (2) and a 19.1\% increase over Model (3). Similarly, for FPG at 8 years, Model (5) achieves a 21.4\% improvement over Model (2) and a 21.7\% increase relative to Model (3). Models (1) and (2) perform similarly, with Model (2) including traditional CGM metrics. These classical models may show greater differentiation with larger sample sizes or additional data.

The full results, including linear term coefficients and other performance metrics such as log-likelihood and UBRE, are available in the Supplemental Material (see Appendix~\ref{sec:append}, Tables~\ref{tab:ta1}--\ref{tab:ta4}).

\section{Discussion}
In this paper, we explore the clinical value of glucose speed and acceleration for long-term glucose prediction using novel functional CGM metrics. Our focus is on two primary clinical outcomes: HbA1c and FPG—key biomarkers for diagnosing diabetes and monitoring disease progression. We utilize baseline data from the AEGIS study as predictors, with outcomes collected five and eight years from baseline, employing a novel functional distributional regression models.

The results demonstrate that incorporating functional glucose fluctuations through marginal distributional representations of speed and acceleration increases the statistical association—in terms of adjusted \( R^{2} \)—by over 20\% compared to using classical CGM and non-CGM metrics alone. This significant improvement underscores the additional time-dependent information captured by these novel CGM metrics. It suggests that CGM can predict diabetes risk in healthy populations and monitor disease progression in diabetic individuals with higher precision than existing biomarkers.

From a practical standpoint, the findings indicate that not just the intensity of glucose, but also the direction (speed) and magnitude (acceleration) of glucose concentration changes are crucial. These metrics, derived from CGM time series, are closely linked to glucose variability \cite{Kovatchev502}—a key factor in dysglycemia and a driver of cellular damage \cite{rolo2006diabetes}. Our findings support the hypothesis that not only hyperglycemic states, but also speed and acceleration at which glucose levels fluctuate, may predict the  deterioration of functional $\beta$-cell mass in Type 2 diabetes. In manifest diabetes, pancreatic $\beta$-cells produce insufficient amounts of insulin to compensate for insulin resistance, resulting in a relative insulin deficiency and leading to hyperglycemia \cite{vskrha2016glucose}.  

While glucose speed and acceleration can be intuitively derived from a time-series of glucose measurements, the physiological traits that govern these dynamics are complex. Glucose speed could closely reflect physiological processes such as glucose absorption and disposal that are in turn dependent on insulin secretion and insulin sensitivity. Glucose acceleration could be related to physiologic modulators of these processes such as incretin secretion and action after meals or the effect of glucagon enhancing endogenous glucose production to decelerate falling glucose after meals.

Our analysis offers a novel perspective on glucose variability by incorporating the glucodensity approach, which includes both marginal and multivariate functional metrics. These advanced metrics provide a deeper understanding of CGM time series beyond traditional measures such as Continuous Overall Net Glycemic Action (CONGA) and Mean Amplitude of Glycemic Excursions (MAGE), which primarily focus on glucose variability without capturing the full dynamics \cite{variabilidadmon, selvin2007short}. We anticipate that future clinical trials will adopt glucose variability as a key outcome measure, moving beyond reliance on average glucose values alone. Our novel CGM metrics can serve as a foundation for developing new quantitative methods in this field.

The traditional marginal glucodensity approach in diabetes research has the clear advantage of simultaneously capturing low, high, and mid-range glucose concentrations. Furthermore, regression models based on these outcomes are highly interpretable. Our previous work on modeling CGM data in clinical trials \cite{matabuena2024multilevel} using the glucodensity approach and multilevel models emphasizes the clinical relevance of this methodology for better understanding og glucose profile evolution during interventions. This provides a more nuanced and sophisticated perspective. In this context, an observational study using glucodensity further highlights the modeling advantages of this approach compared to traditional CGM metrics when comparing  next-generation insulins \cite{gomez2024insulin}.

In another recent study, we demonstrated that the glucodensity approach could predict time-to-diabetes \cite{matabuena2024personalized} (survival model) more accurately than traditional biomarkers, achieving over  13\% improvement in the area under the curve (AUC). Here, we focus on regression modeling because it is a more appropriate statistical method for detecting stronger statistical associations with functional models, especially given the sample sizes we are considering. For this reason, this paper presents multivariate glucodensity analyses primarily for descriptive purposes.

With larger sample sizes, the multivariate glucodensity model—which simultaneously analyzes glucose speed and acceleration—addresses the limitations of marginal models by offering a more comprehensive functional representation. This model has the potential to be more powerful and more interpretable, as evidenced by the  analyses we performed. With a massively increasing use of continuous glucose monitoring in early prevention our approach has the potential to meaningfully extend the predictive capabilities of this diagnostic method.

In conclusion, this work highlights glucose speed and acceleration as novel biomarkers for predicting long-term glucose levels and diabetes outcomes. Our findings emphasize the need for sophisticated analytical tools to leverage CGM data effectively, utilizing functional data analysis methods in digital health solutions \cite{crainiceanu2024functional}.

\subsection*{Funding statement}

This study has been funded by Instituto de Salud Carlos III (ISCIII) through the projects PI11/02219, PI16/01395, PI20/01069 and by the European Union (Programme for Research and Innovation, 2021-2027; Horizon Europe, EIC Pathfinder: 101161509-GLUCOTYPES); and the Network for Research on Chronicity, Primary Care, and Health Promotion, ISCIII, RD24/0005/0010, co-funded by the European Union-NextGenerationEU. The A Estrada Glycation and Inflammation Study (AEGIS) group would like to acknowledge the efforts of the participants and thank them for participating.

\bibliographystyle{plainnat} % or another compatible style
\bibliography{arxiv}         % Replace 'arxiv' with your .bib filename without extension

\appendix

\section{Regression modelling results}\label{sec:append}
\begin{table}[!htbp] \centering 
  \caption{Statistical results for five predictive models to predict the clinical outcome HbA1c at 5 years. The table presents the estimated coefficients for the scalar variables in the linear terms along with their statistical significance. Additionally, we report various measures of predictive capacity and goodness of fit, including Adjusted $R^2$, Log-Likelihood, and UBRE, across the different models.}  
  \label{} 
  \label{tab:ta1}

\begin{tabular}{@{\extracolsep{5pt}}lccccc} 
\\[-1.8ex]\hline 
\hline \\[-1.8ex] 
 & \multicolumn{5}{c}{\textit{Dependent variable:}} \\ 
\cline{2-6} 
\\[-1.8ex] & \multicolumn{5}{c}{HbA1c--5 years} \\ 
\\[-1.8ex] & (1) & (2) & (3) & (4) & (5)\\ 
\hline \\[-1.8ex] 
 age1 & 0.012$^{***}$ & 0.012$^{***}$ & 0.007$^{**}$ & 0.007$^{*}$ & 0.007$^{**}$ \\ 
  & (0.004) & (0.004) & (0.004) & (0.004) & (0.003) \\ 
  & & & & & \\ 
 glu1 & $-$0.002 & $-$0.004 & 0.002 & 0.0002 & $-$0.002 \\ 
  & (0.002) & (0.003) & (0.003) & (0.003) & (0.003) \\ 
  & & & & & \\ 
 AUC &  & $-$0.005 &  &  &  \\ 
  &  & (0.008) &  &  &  \\ 
  & & & & & \\ 
 MAGE &  & 0.006 &  &  &  \\ 
  &  & (0.005) &  &  &  \\ 
  & & & & & \\ 
 CONGA1 &  & 0.002 &  &  &  \\ 
  &  & (0.257) &  &  &  \\ 
  & & & & & \\ 
 Hyper &  & 0.010 &  &  &  \\ 
  &  & (0.008) &  &  &  \\ 
  & & & & & \\ 
 a1c1 & 0.866$^{***}$ & 0.722$^{***}$ & 0.691$^{***}$ & 0.663$^{***}$ & 0.618$^{***}$ \\ 
  & (0.075) & (0.096) & (0.090) & (0.094) & (0.087) \\ 
  & & & & & \\
 Constant & 0.446 & 1.674$^{*}$ & 1.619 & 2.521 & $-$13.997 \\ 
  & (0.301) & (0.893) & (4.083) & (3.404) & (22.819) \\ 
  & & & & & \\ 
\hline \\[-1.8ex] 
Adjusted R$^{2}$ & 0.640 & 0.645 & 0.732 & 0.743 & 0.796 \\ 
Log Likelihood & $-$227.309 & $-$227.634 & $-$202.752 & $-$201.593 & $-$180.094 \\ 
UBRE & 0.511 & 0.513 & 0.407 & 0.405 & 0.333 \\ 
\hline 
\hline \\[-1.8ex] 
\textit{Note:}  & \multicolumn{5}{r}{$^{*}$p$<$0.1; $^{**}$p$<$0.05; $^{***}$p$<$0.01} \\ 
\end{tabular} 
\end{table} 
% Table created by stargazer v.5.2.3 by Marek Hlavac, Social Policy Institute. E-mail: marek.hlavac at gmail.com
% Date and time: Wed, Sep 11, 2024 - 22:22:24

\begin{table}[!htbp] \centering 
  \caption{Statistical results for five predictive models to predict the clinical outcome HbA1c at 8 years. The table presents the estimated coefficients for the scalar variables in the linear terms along with their statistical significance. Additionally, we report various measures of predictive capacity and goodness of fit, including Adjusted $R^2$, Log-Likelihood, and UBRE, across the different models.} 
  \label{tab:ta2}
\begin{tabular}{@{\extracolsep{5pt}}lccccc} 
\\[-1.8ex]\hline 
\hline \\[-1.8ex] 
 & \multicolumn{5}{c}{\textit{Dependent variable:}} \\ 
\cline{2-6} 
\\[-1.8ex] & \multicolumn{5}{c}{HbA1c--8 years} \\ 
\\[-1.8ex] & (1) & (2) & (3) & (4) & (5)\\ 
\hline \\[-1.8ex] 
 age1 & 0.009$^{***}$ & 0.010$^{***}$ & 0.007$^{**}$ & 0.006$^{*}$ & 0.005$^{*}$ \\ 
  & (0.003) & (0.003) & (0.003) & (0.003) & (0.003) \\ 
  & & & & & \\ 
 glu1 & 0.001 & 0.0004 & 0.0003 & $-$0.0002 & $-$0.002 \\ 
  & (0.002) & (0.002) & (0.002) & (0.002) & (0.003) \\ 
  & & & & & \\ 
 AUC &  & $-$0.004 &  &  &  \\ 
  &  & (0.006) &  &  &  \\ 
  & & & & & \\ 
 MAGE &  & 0.002 &  &  &  \\ 
  &  & (0.005) &  &  &  \\ 
  & & & & & \\ 
 CONGA1 &  & $-$0.066 &  &  &  \\ 
  &  & (0.229) &  &  &  \\ 
  & & & & & \\ 
 Hyper &  & 0.006 &  &  &  \\ 
  &  & (0.007) &  &  &  \\ 
  & & & & & \\ 
 a1c1 & 0.699$^{***}$ & 0.647$^{***}$ & 0.695$^{***}$ & 0.702$^{***}$ & 0.724$^{***}$ \\ 
  & (0.062) & (0.093) & (0.096) & (0.098) & (0.099) \\ 
  & & & & & \\ 
 Constant & 1.263$^{***}$ & 1.894$^{**}$ & 1.937$^{*}$ & 1.991$^{***}$ & 1.628 \\ 
  & (0.257) & (0.731) & (1.168) & (0.757) & (1.610) \\ 
  & & & & & \\ 
\hline \\[-1.8ex] 
Adjusted R$^{2}$ & 0.584 & 0.579 & 0.628 & 0.656 & 0.680 \\ 
Log Likelihood & $-$244.277 & $-$247.797 & $-$235.449 & $-$227.939 & $-$222.960 \\ 
UBRE & 0.410 & 0.422 & 0.383 & 0.362 & 0.351 \\ 
\hline 
\hline \\[-1.8ex] 
\textit{Note:}  & \multicolumn{5}{r}{$^{*}$p$<$0.1; $^{**}$p$<$0.05; $^{***}$p$<$0.01} \\ 
\end{tabular} 
\end{table}

\begin{table}[!htbp] \centering 
  \caption{Statistical results for five predictive models to predict the clinical outcome FPG at 5 years. The table presents the estimated coefficients for the scalar variables in the linear terms along with their statistical significance. Additionally, we report various measures of predictive capacity and goodness of fit, including Adjusted $R^2$, Log-Likelihood, and UBRE, across the different models.} 
  \label{tab:ta3}
\begin{tabular}{@{\extracolsep{5pt}}lccccc} 
\\[-1.8ex]\hline 
\hline \\[-1.8ex] 
 & \multicolumn{5}{c}{\textit{Dependent variable:}} \\ 
\cline{2-6} 
\\[-1.8ex] & \multicolumn{5}{c}{FPG--5 years} \\ 
\\[-1.8ex] & (1) & (2) & (3) & (4) & (5)\\ 
\hline \\[-1.8ex] 
 age1 & 0.241$^{***}$ & 0.244$^{***}$ & 0.179$^{**}$ & 0.208$^{**}$ & 0.252$^{***}$ \\ 
  & (0.088) & (0.089) & (0.084) & (0.084) & (0.083) \\ 
  & & & & & \\ 
 glu1 & 0.184$^{**}$ & 0.097 & 0.218$^{***}$ & 0.247$^{***}$ & 0.187$^{**}$ \\ 
  & (0.071) & (0.080) & (0.080) & (0.083) & (0.084) \\ 
  & & & & & \\ 
 AUC &  & 0.140 &  &  &  \\ 
  &  & (0.186) &  &  &  \\ 
  & & & & & \\ 
 MAGE &  & 0.080 &  &  &  \\ 
  &  & (0.152) &  &  &  \\ 
  & & & & & \\ 
 CONGA1 &  & $-$1.030 &  &  &  \\ 
  &  & (7.003) &  &  &  \\ 
  & & & & & \\ 
 Hyper &  & 0.158 &  &  &  \\ 
  &  & (0.211) &  &  &  \\ 
  & & & & & \\ 
 a1c1 & 17.838$^{***}$ & 13.809$^{***}$ & 12.343$^{***}$ & 10.137$^{***}$ & 9.857$^{***}$ \\ 
  & (2.140) & (2.669) & (2.634) & (2.561) & (2.529) \\ 
Adjusted R$^{2}$ & 0.434 & 0.438 & 0.518 & 0.587 & 0.617 \\ 
Log Likelihood & $-$1,859.169 & $-$1,859.422 & $-$1,833.541 & $-$1,809.749 & $-$1,796.796 \\ 
UBRE & 528.949 & 529.761 & 467.466 & 418.508 & 393.725 \\ 
\hline 
\hline \\[-1.8ex] 
\textit{Note:}  & \multicolumn{5}{r}{$^{*}$p$<$0.1; $^{**}$p$<$0.05; $^{***}$p$<$0.01} \\ 
\end{tabular} 
\end{table}

\begin{table}[!htbp] \centering 
  \caption{Statistical results for five predictive models to predict the clinical outcome FPG at 8 years. The table presents the estimated coefficients for the scalar variables in the linear terms along with their statistical significance. Additionally, we report various measures of predictive capacity and goodness of fit, including Adjusted $R^2$, Log-Likelihood, and UBRE, across the different models.} 
  \label{tab:ta4}
\begin{tabular}{@{\extracolsep{5pt}}lccccc} 
\\[-1.8ex]\hline 
\hline \\[-1.8ex] 
 & \multicolumn{5}{c}{\textit{Dependent variable:}} \\ 
\cline{2-6} 
\\[-1.8ex] & \multicolumn{5}{c}{FPG--8 years} \\ 
\\[-1.8ex] & (1) & (2) & (3) & (4) & (5)\\ 
\hline \\[-1.8ex] 
 age1 & 0.320$^{***}$ & 0.388$^{***}$ & 0.286$^{***}$ & 0.323$^{***}$ & 0.218$^{**}$ \\ 
  & (0.103) & (0.099) & (0.099) & (0.104) & (0.110) \\ 
  & & & & & \\ 
 glu1 & 0.355$^{***}$ & 0.270$^{***}$ & 0.277$^{***}$ & 0.252$^{**}$ & 0.463$^{***}$ \\ 
  & (0.097) & (0.098) & (0.095) & (0.101) & (0.113) \\ 
  & & & & & \\ 
 AUC &  & $-$0.394$^{**}$ &  &  &  \\ 
  &  & (0.191) &  &  &  \\ 
  & & & & & \\ 
 MAGE &  & $-$0.439$^{***}$ &  &  &  \\ 
  &  & (0.147) &  &  &  \\ 
  & & & & & \\ 
 CONGA1 &  & 9.980 &  &  &  \\ 
  &  & (7.818) &  &  &  \\ 
  & & & & & \\ 
 Hyper &  & 0.967$^{***}$ &  &  &  \\ 
  &  & (0.237) &  &  &  \\ 
  & & & & & \\ 
 a1c1 & 5.521$^{*}$ & $-$0.567 & 0.366 & 1.187 & 0.228 \\ 
  & (3.067) & (3.766) & (3.872) & (3.903) & (3.903) \\ 
  & & & & & \\ 
Adjusted R$^{2}$ & 0.339 & 0.426 & 0.425 & 0.445 & 0.517 \\ 
Log Likelihood & $-$996.865 & $-$983.198 & $-$983.886 & $-$982.616 & $-$973.459 \\ 
UBRE & 443.177 & 392.443 & 395.010 & 391.652 & 365.831 \\ 
\hline 
\hline \\[-1.8ex] 
\textit{Note:}  & \multicolumn{5}{r}{$^{*}$p$<$0.1; $^{**}$p$<$0.05; $^{***}$p$<$0.01} \\ 
\end{tabular} 
\end{table}
\end{document}